# TITLE PAGE

**Title:** Diversifying the Genomic Data Science Research Community

**Running Title:** Equal Access in Genomic Data Science

**Authors:** The Genomic Data Science Community Network, Rosa Alcazar[1], Maria Alvarez[2], Rachel Arnold[3], Mentewab Ayalew[4], Lyle G. Best[5], Michael C. Campbell[6], Kamal Chowdhury[7], Katherine E. L. Cox[8], Christina Daulton[9], Youping Deng[10], Carla Easter[11], Karla Fuller[12], Shazia Tabassum Hakim[13], Ava M. Hoffman[8], Natalie Kucher[14], Andrew Lee[15], Joslynn Lee[16], Jeffrey T. Leek[8], Robert Meller[17], Loyda B. Méndez[18], Miguel P. Méndez-González[19], Stephen Mosher[14], Michele Nishiguchi[20], Siddharth Pratap[21], Tiffany Rolle[9], Sourav Roy[22], Rachel Saidi[23], Michael C. Schatz[24], Shurjo K. Sen[9], James Sniezek[25], Edu Suarez Martinez[26], Frederick J. Tan[27], Jennifer Vessio[14], Karriem Watson[28], Wendy Westbroek, Joseph Wilcox[30], Carrie Wright[8], Xianfa Xie[31]

[1]Clovis Community College, Fresno, CA, [2]Biology, El Paso Community College, El Paso,TX [3]US Fish and Wildlife (Northwest Indian College), Onalaska, WI, [4]Biology Department, Spelman College, Atlanta, GA, [5]Turtle Mountain Community College, Belcourt, ND, [6]Department of Biological Sciences, University of Southern California, Los Angeles CA, [7]Biology Department, Claflin University, Orangeburg, SC, [8]Department of Biostatistics, Johns Hopkins Bloomberg School of Public Health, Baltimore, MD, [9]National Human Genome Research Institute, National Institutes of Health, Bethesda, MD, [10]Department of Quantitative Health Sciences, University of Hawaii at Manoa, Honolulu, HI, [11]Smithsonian Institute National Museum of Natural History, Washington, DC, [12]Guttman Community College, New York, NY, [13]Department of Microbiology and Biomedical Sciences, Dine College, Tuba City, AZ, [14]Department of Biology, Johns Hopkins University, Baltimore, MD, [15]Department of Biology, Northern Virginia Community College - Alexandria, Alexandria, VA, [16]Department of Chemistry and Biochemistry, Fort Lewis College, Durango, CO, [17]Department of Neurobiology, Morehouse School of Medicine, Atlanta, GA, [18]Science & Technology, Universidad Ana G. Méndez – Carolina, Carolina, PR, [19]Natural Sciences Department, University of Puerto Rico at Aguadilla, Aguadilla, PR, [20]Department of Molecular and Cell Biology, University of California, Merced, Merced, CA, [21]School of Graduate Studies and Research, Meharry Medical College, Nashville, TN, [22]Department of Biological Sciences and Border

Biomedical Research Center, University of Texas at El Paso, El Paso, TX, [23]Department of Math, Statistics, and Data Science, Montgomery College, Rockville, MD, [24]Departments of Biology and Computer Science, Johns Hopkins University, Baltimore, MD, [25]Chemical and Biological Sciences, Montgomery College, Germantown, MD, [26]Department of Biology, University of Puerto Rico – Ponce, Ponce, PR, [27]Department of Embryology, Carnegie Institution, Baltimore, MD, [28]National Institutes of Health, Bethesda, MD, [29]Department of Biology, Flathead Valley Community College, Kalispell, MT, [30]Department of Biology, Nevada State College, Henderson, NV, [31]Department of Biology, Virginia State University, Petersburg, VA

**Corresponding Author and complete contact information:**

Rosa Alcazar

Clovis Community College

Department of Science

10309 N. Willow

Fresno, CA 93730, USA

rosa.alcazar@cloviscollege.edu

Ava M. Hoffman

Bloomberg School of Public Health

Johns Hopkins University

Department of Biostatistics

615 N Wolfe St

Baltimore, MD 21205, USA

ava.hoffman@jhu.edu

Joslynn Lee

Fort Lewis College

Department of Chemistry and Biochemistry

1000 Rim Drive

Durango, CO 81301, USA

jslee@fortlewis.edu

Sourav Roy

University of Texas at El Paso

Department of Biological Sciences and Border Biomedical Research Center

College of Science

500 W University

El Paso, TX 79902, USA

sroy1@utep.edu

# ABSTRACT

Over the last 20 years, the explosion of genomic data collection and the cloud computing revolution have made computational and data science research accessible to anyone with a web browser and an internet connection. However, students at institutions with limited resources have received relatively little exposure to curricula or professional development opportunities that lead to careers in genomic data science. To broaden participation in genomics research, the scientific community needs to support these programs in local education and research at Underserved Institutions (UIs). These include Community Colleges, Historically Black Colleges and Universities, Hispanic-Serving Institutions, and Tribal Colleges and Universities that support ethnically, racially, and socioeconomically underrepresented students in the US. We have formed the Genomic Data Science Community Network (http://www.gdscn.org/) to support students, faculty, and their networks to identify opportunities and broaden access to genomic data science. These opportunities include: expanding access to infrastructure and data, providing UI faculty development opportunities, strengthening collaborations among faculty, recognizing UI teaching and research excellence, fostering student awareness, developing modular and open-source resources, expanding Course-based Undergraduate Research Experiences (CUREs), building curriculum, supporting student professional development and research, and removing financial barriers through funding programs and collaborator support.

# FOUNDATIONS FOR JUSTICE IN GENOMIC DATA SCIENCE

Despite growing opportunities in data science careers, systemic barriers have limited the participation of underrepresented groups in genomic data science research and education (Canner et al. 2017). Among Bachelor's degree recipients in biological sciences, computer sciences, mathematics, and statistics from 2006-2016, 8.7% were Hispanic or Latinx, 7.8% were Black or African American, and 1.9% were multi-racial and/or indigenous American (National Science Foundation 2019a). Meanwhile, these groups represent 16.3%, 12.3%, and 2.5% of the United States resident population, respectively (National Science Foundation 2019a). Disparities are more pronounced in graduate education (Wiley et al. 2020). Affinity organizations where members of underrepresented groups come together are vital to developing a sense of belonging and support system (**Table S1**). However, for true representation in research, science needs inclusive spaces where researchers can communicate actively with educators and students are supported in developing STEM (science, technology, engineering, and mathematics) identities.

The technological advancements of high-throughput sequencing in the last two decades have enabled the rapid proliferation of genomic data (Goodwin et al. 2016) but they have also led to an even greater access imbalance. Over 60 petabases of data (National Center for Biotechnology Information 2021), or about a million times the size of the original human genome project (International Human Genome Sequencing Consortium 2001), are currently available within the US National Center for Biotechnology Information (NCBI) genomic sequencing repositories. This wealth of data will help scientists determine disease risk, diagnose rare conditions, improve drug safety and efficacy (Manolio et al. 2019), survey pathogens for public health applications (Khoury et al. 2020), and even combat the effects of climate change (Hoffmann et al. 2021). Our greatest limitation is personnel to interpret these data. Yet, genomic data science currently lacks a scaffolded mechanism that supports all individuals and provides a hub of intellectual capital, curated genomic data, and the infrastructure required for authentic learning gained through research experiences. Broader, more diverse participation should be the starting point for creating a more inclusive genomic data science field (Mapes et al.

2020). Focusing on participation is not only ethical but desirable for more novel solutions to problems (Hofstra et al. 2020) and necessary for bringing different perspectives to the table (Zook et al. 2017).

Our vision for a diverse scientific community engaged in genomic data science research is one in which researchers, educators, and students thrive in a just and fair system, not limited by their institution's scientific clout, resources, geographical location, or infrastructure (**Figure 1**). Here, we focus on traditionally Underserved Institutions (UIs) in the United States, which include Minority Serving Institutions (MSIs) defined by the US Department of Education: Historically Black Colleges and Universities (HBCUs), Hispanic-Serving Institutions (HSIs), and Tribal Colleges and Universities (TCUs) (Li and Carroll 2007). Underserved Institutions (UIs) also include Community Colleges (CCs) and some Primarily Undergraduate Institutions that overlap substantially with MSIs (Nguyen et al. 2015). Collectively, UIs play a critical role in educating ethnically, racially, and socioeconomically underrepresented students despite limited access to resources (Li and Carroll 2007). In addition to the number of traditionally underrepresented students educated at UIs, these colleges and universities possess unique strengths, such as the greater sense of belonging, more positive mentoring relationships, and role modeling (Gin et al. 2021; Jayabalan et al. 2021). Faculty and staff at these institutions also have experience with the specific interests, needs, challenges, and concerns of the populations they serve (**Text S1**). UIs are therefore essential to removing systemic bottlenecks that lead to a homogenous workforce.

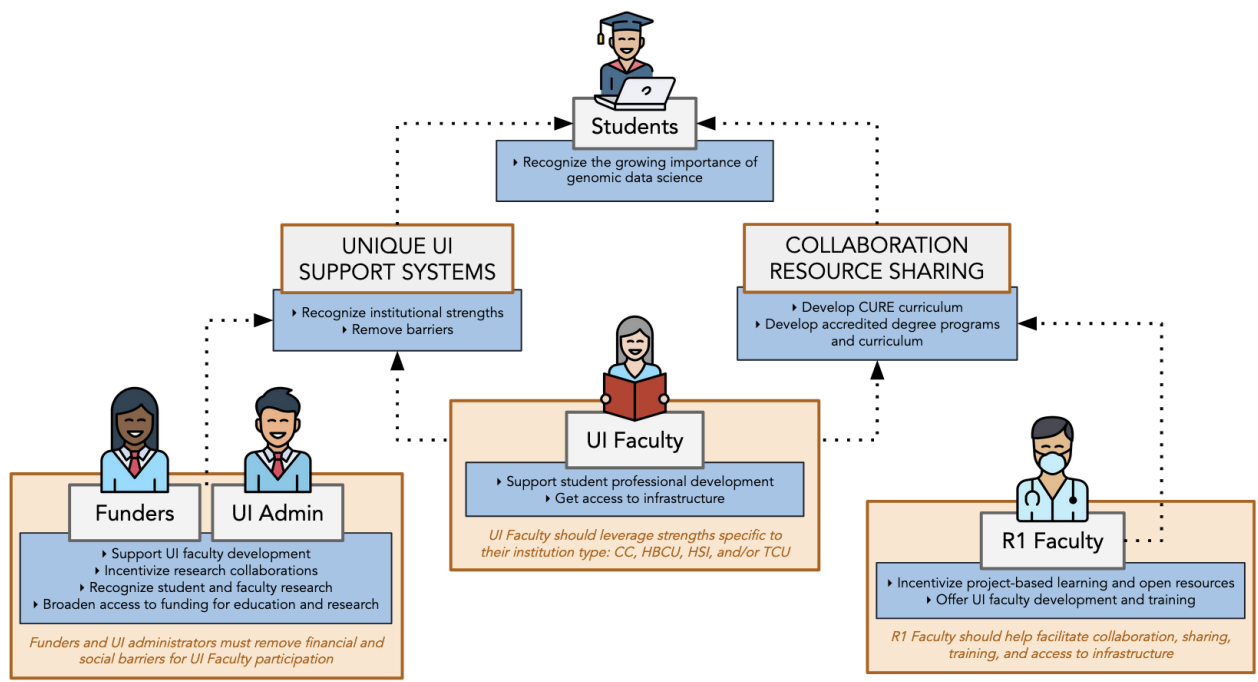

**Figure 1: Our vision for diverse genomic data science.** The Genomic Data Science Community Network strives for a just and fair system where researchers, educators, and students are engaged in genomic data science research regardless of their institution's scientific clout, resources, geographical location, or infrastructure. We propose specific actions (dark/blue boxes) which can be taken by research and educational community stakeholders (light/grey boxes). Support mechanisms are also outlined (italic/orange boxes/hashed lines). UI: Underserved Institution, CC: Community College, HBCU: Historically Black College or University, HSI: Hispanic-Serving Institution, TCU: Tribal College or University.

In this perspective, we share lessons learned from forming the institutionally diverse Genomic Data Science Community Network, GDSCN (http://www.gdscn.org/). Throughout a series of symposiums, one-on-one meetings, and electronic communications, we have provided a platform to listen to the needs of faculty constituents from UIs. Together, we have identified the needs and opportunities in our current academic sphere and its limitations to achieving better representation. While there are several outstanding communities and recent insights for broadening participation in bioinformatics (Jayabalan et al. 2021; Hemming et al. 2019; Buchwald and Dick 2011; Garrison et al. 2019), here we focus on challenges shared broadly by faculty at UIs and action items that can be taken to address them (**Figure 2**). We highlight ways in which outside funders and researchers at R1 institutions – Doctoral Universities with very high research activity (McCormick and Zhao 2005) – can support their colleagues at UIs. With support from funding institutions and a dedicated GDSCN membership, we believe our model will contribute to the organic diversification of genomic data science research.

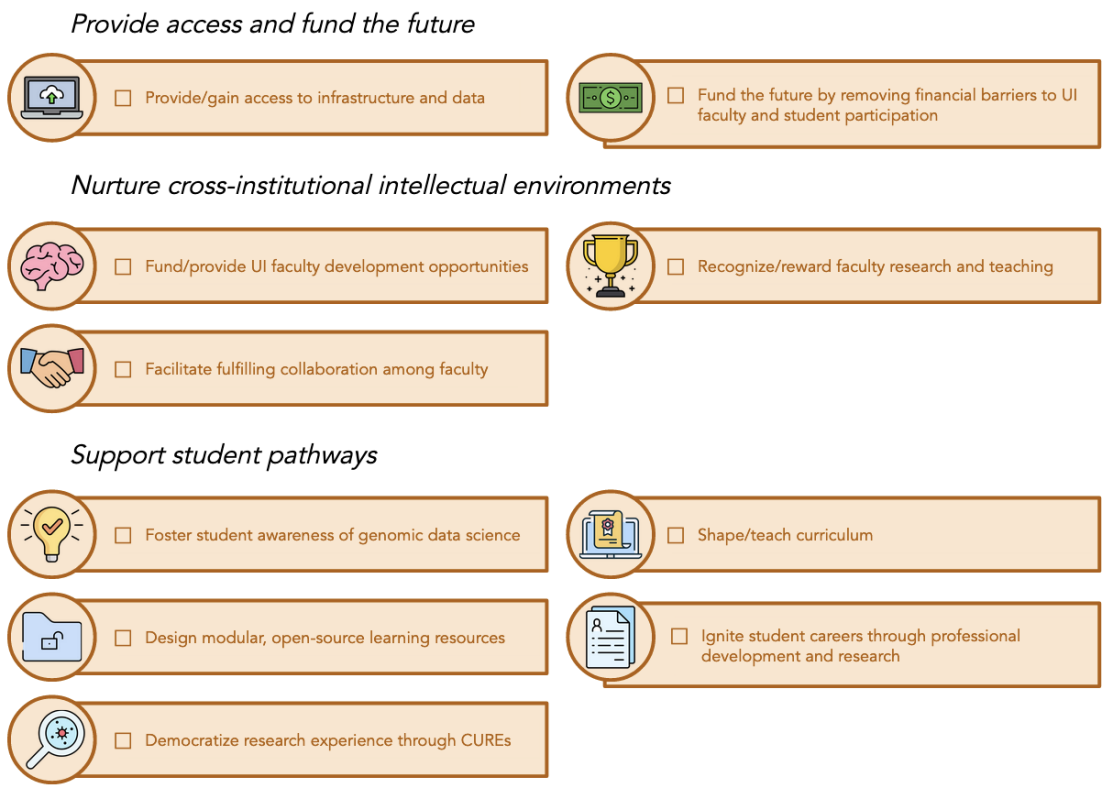

**Figure 2: Checklist of actions for diverse genomic data science.** We envision greater participation in genomic data science with these key actions. UI: Underserved Institution; CURE: Course-based Undergraduate Research Experiences.

# PROVIDE ACCESS TO INFRASTRUCTURE AND DATA

One of the most immediate inequalities is access to data storage and computing power, which are essential prerequisites for genomic data science analyses. Financing computing clusters typically owned and managed by R1 institutions is cost limiting for most UIs, especially if a limited number of faculty members will use it. Cloud computing resources are becoming an increasingly attractive alternative. Institutions only pay for the computing time used, have a lower barrier to entry, require no upfront investment, and ensure students have equal access to software versions (Navale and Bourne 2018) (**Table S2**). For example, the AnVIL platform (Schatz et al. 2021) provides multiple entry points for different levels of experience, from command line access and R/Python notebooks to the no-code-needed Galaxy user interface (Jalili et al. 2020). Fees for compute time and disk storage space are explicitly stated upfront. Other examples include the CyVerse Discovery

Environment (Merchant et al. 2016; Goff et al. 2011) which provides a cloud computing space and includes tools for teaching as well as Galaxy's cloud platform (https://usegalaxy.org/). Laptop or device carts for labs or projects can also reduce the cost burden on students. Combined with affordable devices, such as Chromebooks or tablets, cloud computing can be a powerful, low-barrier resource for democratizing authentic data science experiences in general.

# NURTURE CROSS-INSTITUTIONAL INTELLECTUAL ENVIRONMENTS

## *FUND ONGOING FACULTY DEVELOPMENT*

Broadening participation in genomic data science research depends on a strong support system for faculty at UIs. Yet, resources at these institutions are often stretched thin, making hiring and training the appropriate personnel extremely difficult. Many faculty also have high teaching loads with little time for continuing education, leaving instructors underprepared to teach rapidly evolving genomic data science topics (Williams et al. 2019; Zhan et al. 2019). Financing and incentivizing professional development outreach programs can help meet these challenges. Organizations can follow other successful programs, such as the National Human Genome Research Institute's (NHGRI) Short Course in Genomics (Robbins et al. 2021). This program provides lectures and hands-on learning and has successfully reached over 100 faculty that continue to teach in diverse communities. Events are also hosted by the IDeA Network of Biomedical Research Excellence (INBRE) (National Institutes of Health 2021b). Continued mentorship and resource sharing from NIH researchers and staff after completion of the short course also ensure longer-term positive outcomes.

# *Diversifying Genomic Data Science: Success Stories*

☐ Develop/teach curriculum

I joined the Department of Biological Sciences at St. Mary's University in San Antonio in the Fall of 2018 with a primary mission to develop a Bioinformatics program. There was high demand for a Bioinformatics program at the undergraduate level, with very few Universities offering such a program. We **developed and launched a B.S. in Bioinformatics** with three different tracks in the Spring of 2019 with support from the U.S. Department of Education grant for STEM programs at Hispanic Serving Institutions. I moved from St. Mary's University to the University of Texas at El Paso in the Fall of 2019. However, the program has been running successfully, with **43 students enrolled until the Fall of 2021**. Also, the students from the first batch are **on track to graduate** at the end of the Spring semester, this year.

*- Dr. Sourav Roy, University of Texas at El Paso*

☐ Design modular, open-source learning resources

With funds from NIH, I am developing a module for a **public, free, online genetics course**. The objective of the module is to provide students at tribal colleges a curriculum that **emphasizes Indigenous Genetic research** in relation to (1) sovereignty, (2) stakeholders, and (3) governance. The goals for the student include how to implement both sovereignty and governance in their genetics research with their communities. For this module, interns from the Salish Sea Research Center and I conducted an interview with Lummi Natural Resources hatchery management about the genetics data that is collected on Salmon. The interview questions were proposed by the student interns and reflect the history and the importance of Salmon to the Lummi, as well as the importance of genetics research for hatchery management.

*- Dr. Brandi Cron Kamermans, Northwest Indian College and Salish Sea Research Center*

☐ Facilitate fulfilling collaboration among faculty

The GDSCN has provided a new way of thinking about applying genomics to early college students who lack a coding background. Prior to involvement in the network, I was flailing in my efforts to efficiently develop genomics activities for biology students. Plugging into the resources of NIH and JHU, while **connecting with like-minded community college colleagues**, has saved countless hours while developing concrete programs for community college students that can be used with little modification. A surprising side benefit has extended beyond the United States. It has been encouraging to introduce our work to European medical schools who instantly recognize the necessity of applying genomics study to their early medical students. Beyond the practical benefits of **connecting with experts** from across the nation through GDSCN, it has been a tremendously **reaffirming, supportive, and encouraging** group for advancing genomics teaching.

*- Andrew Lee, Northern Virginia Community College*

☐ Democratize research experience through CUREs

I work to create pathways for every student, no matter their background, socioeconomics, or clout of their home institution. Working at Clovis Community College allows me to help diverse populations reach academic goals that are otherwise impossible. These students are often first-generation college students and/or from minority groups underrepresented in STEM. My collaborators and I are developing a miniCURE, short modules introducing genomics data science to freshmen with a focus on the scientific process. The goal of the miniCURE is to **provide access and exposure** to the field and its power to transform biomedical research. We aim to be fun, work collaboratively, and create a **sense of belonging** in STEM. The capstone is a poster presentation at a local or regional symposium. The miniCURE has been offered for **8 semesters and 300+ students** by two different instructors. Students express motivation to take higher level courses and several students have expressed that "Things didn't click in the Biology for Majors Course until this project".

*- Dr. Rosa Alcazar, Clovis Community College*

☐ Democratize research experience through CUREs

Trainings in genomics and bioinformatics are keenly needed at Historically Black Colleges or Universities (HBCUs). While the **COVID** pandemic has disrupted the normal educational and research activities, it has provided a great opportunity to teach students about **genomics, bioinformatics, and evolution**. Seizing this unique opportunity, I developed a research project for students who worked with me for their capstone Investigations and Research course to study how the **SARS-CoV-2 virus** genomes have been evolving and how new variants could emerge due to recombination using publically available genomic sequences of the virus and innovative bioinformatic analysis methods, which has led to a **manuscript** available in public domain and is still ongoing. In collaboration with colleagues at mathematics and computer science, we have also established an interdisciplinary genomics and bioinformatics program at the Virginia State University, a HBCU.

*- Dr. Xianfa Xie, Virginia State University*

**Figure 3: Success stories informing the right paths forward.** GDSCN members provided non-exhaustive examples of successes at their institutions. These examples demonstrate how UI faculty can excel in nurturing collaborative environments and supporting student pathways with the right support network.

Outside of government sponsored programs, small grants funding workshops, conferences, or assistants (such as research or teaching assistants) can substantially empower UI faculty. Researchers at R1 institutions can build on-demand materials for self-directed faculty education and/or tailor "train the trainer" workshops for faculty needs and interests. For example, the Jackson Laboratory designed a week-long course, Big Genomic

Data Skills Training for Professors, to train small college and regional university faculty (Zhan et al. 2019). Leaders at R1 institutions should commit to honoring these contributions when considering teaching, service, outreach, and/or tenure committee decisions.

Within UIs, administrators should encourage protected time and teaching release for attending workshops, writing proposals, and/or developing new curriculum. They should incentivize faculty members to pursue educational and research opportunities in genomic data science for the students and the university. Dedicated sabbatical time allows UI faculty to develop research and coursework with outside faculty (Hemming et al. 2019) and innovate new student experiences (Yarmohammadian et al. 2018). Future initiatives can capitalize on a sabbatical internship/exchange model, where faculty can choose to travel to different research labs and gain exposure to additional research initiatives and peer mentorship. Not only do these professional experiences enhance skill-building, they also help future collaborators find one another and organically build collegiate networks.

Faculty members at UIs who have research expertise should be incentivized to build up their colleagues through peer sharing and learning. Programs like the Centers of Research Excellence in Science and Technology (CREST) and HBCU Research Infrastructure for Science and Engineering (HBCU-RISE) (National Science Foundation 2017a), as well as INBRE specifically support these efforts within UIs. Additionally, the recently established NHGRI Office of Training, Diversity and Health Equity (TiDHE) aims to coordinate training and career development for a more diverse genomics workforce (National Institutes of Health 2021c). Faculty who are entirely dedicated to teaching can use these funds to engage students in authentic research activities and to explore pedagogical studies. Furthermore, with administrative support within the same institution, experienced faculty can help develop "pathways"(from beginner to more advanced) by which their colleagues can progressively learn (Crown et al. 2011). Going beyond one-off workshop experiences to form development plans could help make faculty participation both effective and scalable.

## FACILITATE FULFILLING COLLABORATION

Collaboration among institutions ensures faculty come in contact with a diversity of topics and perspectives and are set up for success, from proposal to publication (Hemming et al. 2019). All UI faculty should consider meeting collaborators through established programs like the National Research Mentoring Network (NRMN, https://nrmnet.net/) (Hemming et al. 2019), which connects experienced research mentors with student and faculty mentees from Minority Serving Institutions. The Oak Ridge Institute for Science and Education (ORISE) matches faculty with mentors for hands-on research experience at one of the US Department of Energy's sponsoring agencies (https://orise.orau.gov/). Affinity groups, such as the Native Investigator Development Program (NIDP) and others (**Table S1**), can also contribute toward fruitful, collaborative relationships (Buchwald and Dick 2011). Government programs can also help sponsor additional community building. For example, the Research Coordination Networks Program aims to foster synthesis, new collaborations, and resource sharing while advancing research and education (National Science Foundation 2017b).

Collaborative networks must position UI faculty in leadership roles while actively engaging them in setting goals and priorities. Currently, only a small number of programs specifically target these faculty. For example, the INBRE and Bridges to the Baccalaureate (National Institutes of Health 2021a) programs sponsor collaboration between 2-year and 4-year UI teaching institutions and R1 research institutions with the goal of supporting faculty and providing student research opportunities. Other collaborative networks, such as the Network for Integrating Bioinformatics into Life Sciences Education (NIBLSE), aim to support faculty in bioinformatics coursework while encouraging professional achievement via publication (Dinsdale et al. 2015). In addition to connecting through scientific conferences and meetings, hands-on workshops for faculty and/or students, bioinformatics outreach efforts, and seminar series, these programs can help move toward balancing access to collaborative networks.

## RECOGNIZE FACULTY RESEARCH AND TEACHING

Recognizing the research breakthroughs by faculty has a two-fold benefit: it rewards faculty members' hard work and raises awareness and discussion in the broader community. Sharing the research experiences on social media platforms, such as Twitter (https://twitter.com/), is perhaps the fastest and easiest way to gain academic recognition and raise awareness (Cheplygina et al. 2020). Faculty should work together to amplify each other's work. Social media-savvy faculty might consider social media workshops for those who are not used to engaging on these platforms. Where possible, faculty should leverage their institutional press/media office to create releases and/or posts or use organizations that aggregate blog or social media posts. Using social media and blogs to disseminate work can raise important questions, such as when and how work should be shared among stakeholders, from public relations offices at institutions to Tribal leadership. Research symposia can also serve to showcase faculty participation and collaboration while making them feel included in the community. Existing programs, such as the NHGRI Short Course in Genomics (Robbins et al. 2021) provide faculty the opportunity to present work to NIH teams as well as other conference communities. Virtual conferences, having evolved quickly during the COVID-19 pandemic (Jarvis et al. 2020), can provide both a low-cost and accessible way to feature faculty and student research experiences.

Faculty need to be acknowledged for their excellence in research as well as their hard work teaching and applying for funding, regardless of award status. Non-financial awards at all levels, from institution and school or division to department, are relatively easy to establish for innovative contributions to curriculum content, teaching, research, service, outreach, and mentoring. Administrators can show faculty they are valued by considering such awards in tenure or promotion decisions. As courses in genomic data science are often co-taught by faculty members from different disciplines or with different expertises, university administrators should also create straightforward methods to attribute teaching credits to each participating faculty member and count that in their total teaching load. Overall, empowering UI faculty will lead to greater diversity among faculty researchers in genomic data science (Whittaker et al. 2015). Improved faculty representation also paves the way for improved student achievement, recruitment, and retention (Merisotis and McCarthy 2005).

# ENGINEER MULTIPLE STUDENT PATHWAYS

## *FOSTER STUDENT AWARENESS*

With infrastructure and a faculty foundation established, engaging students and the community directly is required for broadening long-term student participation. In fact, a recent survey highlights that a tribal community has specifically requested more genetics education within their community (Claw et al. 2021). Based on our assessment of the role of CCs, HBCUs, HSIs, and TCUs within their communities, these are the best institutions to provide this education. However, students starting their post-secondary education with interests in biological sciences might have little to no exposure to genomic data science as a potential area of study and career path. Moreover, the sheer volume of topics to cover in introductory biology courses and time needed for molecular "wet" lab instruction can make incorporating additional bioinformatics and genomics topics difficult. Existing resources also tend to be less specialized. Thus, to facilitate early student exposure to genomic data science, faculty likely need to restructure and/or update introductory biology courses to include essential material on genomics and bioinformatics.

To foster early awareness, instructors can briefly introduce genomic data science topics in first-year/freshman colloquia (e.g., (Clark and Cundiff 2011), seminars, and professional/career development courses. Topics should highlight the challenging and fulfilling career opportunities and the potential for a meaningful impact within their communities. Technical development and bioethics should also be emphasized in these courses. Diverse interests, such as computer science, statistics, or other interdisciplinary majors, can be accommodated via course cross-listing. Departments can also host joint seminar series where students have the opportunity to meet genomic data science researchers working in academic, industrial, and/or nonprofit careers. College administrators should support faculty as they develop best practices for introducing this rapidly growing discipline to students.

The final years of secondary school are a pivotal time for students. Faculty should focus on raising awareness, including via open houses, easy-to-find websites, workshops, and weekend learning experiences. Reaching out to high school student STEM clubs, or sponsoring the development of such clubs, can serve as a channel by which to encourage participation. Providing opportunities for college or university visits for students can improve student self-efficacy and give them the tools they need to select their postsecondary career path (Glessner et al. 2017; Boatright 2020). A one-day congress that brings students from multiple high schools to a college or university could allow students to meet practicing scientists. Early College High Schools (ECHS) and Dual-enrollment programs, which allow students to simultaneously earn a high school diploma and an Associate's degree, could provide a conduit for reaching students historically underrepresented in higher education (Texas Education Agency 2020; Virginia Community Colleges 2020). Development of strong partnerships between the high school independent districts and higher education institutions is key to the success of these initiatives. The DataTrail (https://www.datatrail.org/) program also provides a successful model by which high schoolers and GED earners can connect directly to faculty mentors and data science employment.

Raising awareness among even younger students in primary and secondary school is a recognized need (Morgan et al. 2016). Fundamental education in programming (e.g., via games, (Lindberg et al. 2019)) can lay the foundation for scientific analysis; the genomic data science community should work with teachers to provide age-appropriate activities. Access to computers is a consistent challenge, although smart devices like phones and tablets can be powerful tools when combined with cloud resources (**Table S2**). Despite technological advancements, preschool and kindergarten teachers often feel unsupported and underprepared to teach STEM, contributing to underrepresentation that is persistent and pervasive (Fuller et al. 2021). Remedying resource and training inequities must be a funding priority as government agencies consider broader initiatives to introduce data science more prominently into primary and secondary school education.

## DESIGN MODULAR, OPEN-SOURCE LEARNING RESOURCES

Most life science instructors agree that bioinformatics knowledge and skills are essential to biology students (Wilson Sayres et al. 2018). One of the great successes of data science more broadly is the proliferation of open source and open access resources. Many freely available online resources have emerged with the growing interest in genomic data science, including the Coursera Genomic Data Science Specialization (https://www.coursera.org/specializations/genomic-data-science) and Bioinformatics Algorithms (https://www.bioinformaticsalgorithms.org/). Both of these courses have already reached hundreds of thousands of students. While massive online open courses (MOOCs) are broadening access to education, additional support (such as community building and awareness, confidence building, faculty training) is needed for learners from underrepresented groups (Stich and Reeves 2017). UI faculty can serve as important bridges between MOOC material and UI students, but instructors still face many ongoing challenges. UI faculty must wade through the wealth of material to find content that is appropriate and interesting for their target audience or is specific to a particular application, topic, or technique. Finding activities that are modular and interchangeable given their existing course lesson plans takes time. In some cases, restrictive licensing for "remixing" and updating as needed is a barrier. Many educators, UI and R1 alike, also struggle to keep pace with constantly evolving approaches (Ryder et al. 2020).

Because many resources already exist to teach data science more generally, new development can focus on free interactive platforms (Kross et al. 2020) and/or open source methods for updating and remixing content to make it more accessible (Savonen et al. 2022). In general, resources teaching introductory concepts, foundational statistics, and programming skills should continue to be made open source wherever possible (**Table S3**). For more specific genomics content, R1 developers should work with UI faculty to develop relevant, approachable content that can be freely accessed and tailored as needed. These partnerships can be incentivized by administration and sponsoring agencies by including it in promotion and/or funding decisions. Another successful approach is to include instructor guides within materials, such as the teaching instructions included with the modular Microbiome QIIME2 activity developed by UI faculty (see GitHub link: Lee 2019).

## DEMOCRATIZE RESEARCH EXPERIENCE THROUGH CUREs

Undergraduate research experiences can improve outcomes for historically underrepresented students in the sciences (Carpi et al. 2017; Krim et al. 2019). Course-based Undergraduate Research Experience, or CUREs, bring students directly to authentic research with their peers in a classroom setting and are defined by use of scientific practices, discovery, relevant and/or important work, collaboration, and iteration (Auchincloss et al. 2014). Whereas government-focused programs (such as the NSF REU program) traditionally offer limited lab space to a small number of students, CUREs expose a larger group of students in tandem (Elgin et al. 2021). CUREs break down student financial and personal barriers and can minimize faculty members' unconscious bias (Bangera and Brownell 2014). Furthermore, flexible CUREs can help establish a mutually uplifting, safe, and positive mentoring environment while creating a peer community where students can collaborate, share discoveries, and support one another (Kokotsaki et al. 2016). Indeed, a sense of peer and institutional belonging is a key metric for student feelings of inclusion and retention in UIs (Winkle-Wagner and McCoy 2018; Sweeder et al. 2021).

Though CUREs have many benefits (Elgin et al. 2021), UI faculty sometimes receive pushback for implementing CUREs or requesting research classroom hours within established curricula. One solution is creating a variety of CUREs with flexible length / timing (weeks or semester-long) in mind. For example, some instructors might find it easier to incorporate "mini-CUREs" that require only part of the semester within an existing course. Faculty can also  consider incorporating CUREs within a senior capstone course. It is essential that faculty collectively maintain a support system for sustainable CURE implementation (Lopatto et al. 2014; Elgin et al. 2021). Research faculty at R1 institutions can contribute to this support system by helping develop CUREs, making them more accessible, and/or providing "train the trainer" resources to UI faculty colleagues. The Genomics Education Partnership (https://thegep.org/) provides training and resources for faculty implementing CUREs. Programs like the Small World Initiative (a student education program focused on widespread antibiotics resistance, http://www.smallworldinitiative.org/) have trained faculty at over 300 schools to incorporate authentic research into their classrooms.

The key for success for any classroom content is to ensure it is relevant to real life and/or the students' future career goals and that it is an authentic research experience for both the faculty and students involved. There are several examples of CUREs that have been created by UI faculty and/or implemented at UIs. The SEA-PHAGES (Science Education Alliance-Phage Hunters Advancing Genomics and Evolutionary Science, https://seaphages.org/) network is both an inclusive research education community and the developer of the two-term laboratory CURE studying viral diversity, evolution, genome annotation, and bioinformatic analyses (Jordan et al. 2014; Hanauer et al. 2017). Data generated by SEA-PHAGES students have been used in real-time medicinal interventions and high-impact publications (Dedrick et al. 2019), which can leave a lasting impact on students. The Clovis Community College Biol-12 Research in Biology course (https://www.cloviscollege.edu/landing/biol-12-genomics-data-science.html) introduces data science with a unique emphasis on university transferability while having no prerequisites (**Figure 3**). This course also includes professional development for students, including presenting at a symposium and working on college or job applications. With funding from the NIH Bridges to the Baccalaureate Program, faculty at El Paso Community College developed Introductory Biology CURE projects focused on topics relevant to the local US-Mexico Border community including water contamination and antibiotic resistant microorganisms. Faculty at Virginia State University designed a research program focused on SARS-CoV-2 for the required one-semester capstone course (**Figure 3**). This work demonstrated the real scientific impact of students and faculty to help reveal a biological process that has been largely overlooked by existing studies of the SARS-CoV-2 virus (Xie et al. 2021). Development of new CURE topics, for example DNA forensics, population genetics of communities, or bioinformatics in nature, should thoughtfully engage with students throughout the design process.

## *SHAPE CURRICULUM*

Accreditations for students participating in genomic data science courses and CUREs are critical for state-of-the-art education, retention, and enrollment growth. Yet, building new degree programs is challenging, especially for broad, rapidly evolving domains like genomic data science. Course names like "Big Data" might mean very different content, activities, platforms, and learning objectives among institutions. First, programs should employ a "pathway model" – consisting of clear program outlines/degree plans with distinct course sequences, learning objectives, progress milestones, and learning outcomes. Setting up guidelines and learning objectives that would need to be covered early on will also help guide individual courses. Standardizing learning outcomes and course titles (e.g., topic + tool, "Genomic Data Science with Galaxy") will provide a more detailed picture of the student's exposure to specific skills and topics. To allow students to explore tools more in depth, some institutions might find it beneficial to include coding languages (e.g., R or python) as part of the computer literacy or general education requirements.

Faculty at UIs have already been successful in piloting genomic data science curriculum. The St. Mary's University B.S. in Bioinformatics (**Table S4**) was launched in the Fall of 2019, and currently has 43 students enrolled in the program (**Figure 3**). The Genomics and Bioinformatics Program at Virginia State University provides training in genomics, computer science, and statistics for students from different majors including biology, computer science, mathematics, and agriculture. These students take the same high-level genomics and programming for bioinformatics courses during their third year and conduct research throughout their degree, culminating in a one-year extensive research course during their senior year.

Partnerships among research and undergraduate-focused institutions can be a powerful tool for modeling transferability and curriculum standardization. For example, New Jersey's 3+1 program (Rowan College at Burlington County 2021), University of California-Los Angeles' Center for Community College Partnerships (https://www.aap.ucla.edu/units/cccp/), and Virginia's Statewide Transfer Agreement (https://www.vccs.edu/transfer-programs/) partner community colleges with four-year research institutions, reducing the cost burden on students seeking genomic data science degrees. Students obtaining an associate degree and earning a minimum grade point average from Virginia's community colleges can guarantee admission to any of more than 30 Virginia colleges and universities, including those offering B.S degrees in Bioinformatics or Data Science. Faculty and administrators involved with these degree programs at R1 institutions can better support incoming students by offering tailored recommendations, such as suggested prerequisites. This translates to a broader diversity of transfer students set up for success.

## IGNITE CAREERS THROUGH DEVELOPMENT AND RESEARCH

Data science careers have grown rapidly in the past decade, providing an exciting opportunity as companies are working hard to fill the global talent gap (Rawlings-Goss 2019). However, students might not be well acquainted with the next steps in their professional career beyond academic credit. Mentorship to help students explore and/or identify further career opportunities would help bridge this gap, whether students are applying to summer research programs, transferring to research institutions, exploring graduate school degrees, or entering the workforce. To remove barriers for applying to summer research programs and other positions, R1 and UIs with larger research programs should create application guides, offer workshops, information sessions, recruiting events, and/or alumni panels. Visual tools for mapping opportunities by location, topic, and experience level (i.e., undergraduate or graduate) could also be useful. Next-step opportunities offered by research institutions should provide clear information about stipends and flexibility, if possible (Jensen et al. 2021). Creating new student-focused symposia, even if small in size, is another straightforward way to engage a broad array of students by building confidence and creating a supportive and engaging experience among peers. Funding agencies and research institutions should be incentivized to make introductions and help build broader networks that include industry, non-profit, and other non-academic data scientists. Targeted training for different groups can also be an excellent way to help students transition to next steps in their careers. For example, the Summer internship for INdigenous peoples in Genomics (SING, https://sing.igb.illinois.edu/) workshop trains indigenous peoples in the concepts and methods currently used in genomic research while discussing the uses, misuses and limitations of genomics as a tool for indigenous peoples' communities. Finally, students could also be encouraged to consider outside accreditation, such as the existing American Statistical Association Accreditation (https://www.amstat.org/ASA/Your-Career/Accreditation.aspx) to demonstrate their achievements where internal certificates or programs are unavailable.

From the student perspective, identifying outside learning and professional development requires knowing what and where to search ahead of time. This can be remedied by engaging students earlier during secondary education. While college visits are useful for building interest in postsecondary education in general, a one-day science congress targeted toward high school students hosted by a college or university could create connections between students and practicing scientists. These connections could help students find opportunities and could even evolve to closer mentor-mentee relationships. For students already at 2-year or 4-year teaching institutions, faculty and students would benefit from attending events like the Annual Biomedical Research Conference for Minority Students (https://www.abrcms.org/) or the Society for Advancement of Chicanos/Hispanics and Native Americans in Science annual meeting (https://www.sacnas.org/), which provide pathways for further experiences and networking.

## FUND THE FUTURE

Faculty members at UIs have been historically disadvantaged when competing for funding due to the high teaching load and/or perceived lower research quality, lack of research infrastructure and research personnel, and stigma associated with reanalyzing existing data, among other reasons (Ginther et al. 2011; Hemming et al. 2019). Systemic biases also exist due to a lack of understanding of the importance of institutions that enroll significant numbers of traditionally underrepresented students. While several pivotal funding opportunities for UIs have emerged in recent years, the GDSCN constituency identified where funding agencies and philanthropies can prioritize resources. These include funds for faculty networking and knowledge exchange, CURE development, and cloud computing, alongside ongoing efforts to strengthen research support.

Events where faculty can network and develop collaborations require minimal funds, but are still a major need among UI faculty (Hemming et al. 2019). Such events have long lasting benefits, as they facilitate novel research, educational materials, and discovery. Faculty at R1 research institutions can also commit to contacting UI faculty to submit grant proposals collaboratively, share open access teaching resources, and/or invite students to participate in external CUREs and other research experiences. Programs can also be enacted to fund R1 faculty research or teaching time at UIs. Going forward, evaluation of larger NIH proposals could incorporate a Broader Impacts section akin to those required by NSF, where R1 researchers regularly commit to engaging and collaborating with underrepresented communities. Modest funding could also make a sizable difference for recruiting outside speakers. Speakers coming from underrepresented groups and/or with non-traditional backgrounds can often connect more effectively with students of similar backgrounds and should be encouraged to do so with financial compensation. Grants supporting potential role models could also be expanded by government agencies, for-profit companies, and philanthropies. Postdoctoral and early career fellowships are particularly lacking for underrepresented groups.

As mentioned above, there is a much deeper root for the underrepresentation of certain communities in the field of genomics, which may go back to the scientific education they receive in early K-12 levels, besides socioeconomic and family factors. Therefore, to increase the participation of the currently underrepresented communities in the field of genomics, there needs to be greater investment in science and mathematics at the K-12 grade levels and more activities engaging K-12 students in understanding biology, genomics, and related sciences. Besides improving the overall education quality and particularly science and mathematics education in grade schools, federal funding agencies and philanthropies should encourage faculty with expertise in genomics working at either UIs or R1 universities to provide training to science and mathematics teachers at high schools, secondary schools, and even primary schools and allow financial support to do so.

The NIH and NSF provide grants for curriculum development, such as the NIH Research Enhancement Award for small-scale research projects at educational institutions (National Institutes of Health 2021d), the Advancing Innovation and Impact in Undergraduate STEM Education at Two-year Institutions of Higher Education program (National Science Foundation 2021), and the Historically Black Colleges and Universities - Undergraduate Program (National Science Foundation 2020). Yet, the percentage of funding granted to UIs remains very low. Releasing teaching time also requires other faculty to shoulder the released teaching load, making it difficult to implement these programs without significant funding. Similarly, the successful NSF REU program (National Science Foundation 2019b) also supports relatively few students, typically only ~10 students per site, which is orders of magnitude smaller than is necessary to give all students an opportunity. Funding programs that sponsor scalable and accessible CUREs can likely provide greater collective impact (Elgin et al. 2021). Philanthropies and private companies could also consider partnering with UI faculty to fund CURE modules in specific topic areas.

More generally, funders must support research in underrepresented communities and at UIs. UIs are typically under much higher economic pressure, making it difficult for faculty to get started with research. First, funders must consider computing costs as part of the solution. Options include streamlining funding for cloud computing through programs like NIH Strides and public-private partnerships like the NSF CISE-MSI program (NSF Directorate for Computer and Information Science and Engineering with support from Google) (**Table S2**). Importantly, platforms should consider a free no-strings-attached tier for users where compute hours regenerate over time. This prevents accidental overspend, makes it easier to get started, and makes practicing with students simple. While government agencies have been increasing support for UI research in recent years, companies and private philanthropies can often adopt change more quickly. For example, the Chan Zuckerberg Initiative (CZI) Essential Open Source Software program (https://chanzuckerberg.com/rfa/essential-open-source-software-for-science/) aims to improve reproducibility and transparency in biology and medicine while explicitly seeking applications that increase diversity and inclusion. Biotechnology companies can work directly with UIs to offer support and a path to employment for students. Other research challenges require creative solutions. For example, UIs often lack institutional journal access, which could be overcome with sponsorship by R1 institution libraries. Sponsoring staff positions can also work toward reducing the administrative burden on UI faculty. Funders will need to continue to engage their target communities to provide usable and meaningful support. The GDSCN has created a growing database of resources (https://www.gdscn.org/resources) as a starting point for bringing coursework, development, and funding resources to UIs and their communities.

# THE GENOMIC DATA SCIENCE COMMUNITY NETWORK VISION

We propose a vision where researchers, educators, and students from diverse backgrounds are able to fully participate in genomic data science research. Underserved Institutions (UIs) make education more accessible to students from underrepresented groups. Our strategy is to provide resources that are currently absent at these key institutions to augment the unique strengths that these institutions already provide. As a first step towards this vision, we planted the seeds for a Genomic Data Science Community Network (GDSCN) by bringing together faculty from CCs, HBCUs, HSIs, and TCUs alongside faculty and staff from Johns Hopkins University and NHGRI. This vision and recommendations in this manuscript represent the culmination of a year-long process that involved one-on-one virtual meetings, four virtual synchronous events along with several rounds of asynchronous writing sessions.

In addition to bringing faculty together, the pilot efforts of the GDSCN has also supported tangible resources. First, we are developing several modules of genomic data science curricula specifically targeted toward UI student audiences, available at https://www.gdscn.org/curricula/courses. Through NHGRI, GDSCN and other UI faculty members are provided access to cloud computing credits on the AnVIL platform in order to run these courses in their classrooms. Simultaneously, we organized a series of symposiums, where we crowdsourced challenges, needs, and opportunities in the current academic infrastructure as well as barriers that need to be overcome to enable broader participation and inclusion in genomic data science research. Finally, we initiated the development of several modules within the Open Case Studies Project (https://www.opencasestudies.org/). In the future, we hope to grow the network (https://www.gdscn.org/contact-us) to incorporate more perspectives and support more faculty from UIs to meet their networking, research, and education goals. We plan to continue iterating on our initiatives via GDSCN faculty surveys. We will likewise record the number of faculty and students reached in training events, classroom time, and upcoming events (such as grant writing workshops).

More broadly, the National Institutes of Health (NIH) has identified four major underrepresented groups whose participation is needed to increase diversity in biomedical research: racial and ethnic groups, people with disabilities, people with low socioeconomic status, and women (National Institutes of Health 2019). In this perspective, we focus on UIs, including Minority Serving Institutions. Our current scope does not specifically include people with disabilities and women. However, both of these are identities that intersect with UIs. We expect that the actions discussed here (**Figure 2**), such as greater access to CUREs, exploration of relevant research topics and issues, and expanded funding, will be relevant to other underrepresented communities. However, the GDSCN will need to have direct engagement with women and people with disabilities in the genomic data science community to determine the best way to leverage their specific strengths and advocate for their most pressing needs.

In the future, we hope that faculty at R1 institutions—who identify with an underrepresented group or share an interest in improving the education and research in genomic data science at the institutions serving these students—will consider lending their perspective to the community and reach out for collaborations with the faculty and students at these institutions. Ultimately, we are hopeful that these collaborations will empower scientists to solve key problems, such as the pervasive bias in genomic data collection (Sirugo et al. 2019) and ongoing challenges with accountability, data sovereignty, community/tribal consent, and data misuse (Hudson et al. 2020). Although we have limited our initial efforts to institutions in the US, we believe that lessons learned among the GDSCN will spark discussions in the global research community. We also believe that GDSCN efforts will be improved by including student and tribal representatives in the conversation (Claw et al. 2021). Finally, we hope that our insights and materials will encourage scientists and their teams to learn and actively support UI communities, ultimately improving access, removing barriers, and making science more innovative and inclusive as a whole.

# COMPETING INTERESTS STATEMENT

The authors declare no competing interests.

# ACKNOWLEDGEMENTS

This work is dedicated to the late James Peter Taylor, the Ralph S. O'Connor Professor of Biology and Computer Science at Johns Hopkins University, who was one of the original architects of the AnVIL and an ardent champion for open science (https://galaxyproject.org/jxtx). We also thank B. C. Kamermans for helpful feedback. The GDSCN is supported through a contract to Johns Hopkins University (75N92020P00235). The AnVIL is supported through cooperative agreement awards from NHGRI with co-funding from OD/ODSS to the Broad Institute (#U24HG010262) and Johns Hopkins University (#U24HG010263).

# REFERENCES

Auchincloss LC, Laursen SL, Branchaw JL, Eagan K, Graham M, Hanauer DI, Lawrie G, McLinn CM, Pelaez N, Rowland S, et al. 2014. Assessment of course-based undergraduate research experiences: a meeting report. *CBE Life Sci Educ* **13**: 29–40.

Bangera G, Brownell SE. 2014. Course-based undergraduate research experiences can make scientific research more inclusive. *CBE Life Sci Educ* **13**: 602–606.

Boatright T. 2020. The Impact of the College Tour Element Embedded in a High-quality Afterschool Program. ed. M. Morris., Northeastern University, Ann Arbor, United States https://www.proquest.com/dissertations-theses/impact-college-tour-element-embedded-high-quality/docview/2395333563/se-2.

Buchwald D, Dick RW. 2011. Weaving the native web: using social network analysis to demonstrate the value of a minority career development program. *Acad Med* **86**: 778–786.

Canner JE, McEligot AJ, Pérez M-E, Qian L, Zhang X. 2017. Enhancing Diversity in Biomedical Data Science. *Ethn Dis* **27**: 107–116.

Carpi A, Ronan DM, Falconer HM, Lents NH. 2017. Cultivating minority scientists: Undergraduate research increases self-efficacy and career ambitions for underrepresented students in STEM. *J Res Sci Teach* **54**: 169–194.

Cheplygina V, Hermans F, Albers C, Bielczyk N, Smeets I. 2020. Ten simple rules for getting started on Twitter as a scientist. *PLoS Comput Biol* **16**: e1007513.

Clark MH, Cundiff NL. 2011. Assessing the Effectiveness of a College Freshman Seminar Using Propensity

Score Adjustments. *Res High Educ* **52**: 616–639.

Claw KG, Dundas N, Parrish MS, Begay RL, Teller TL, Garrison NA, Sage F. 2021. Perspectives on Genetic Research: Results From a Survey of Navajo Community Members. *Front Genet* **12**: 734529.

Crown SW, Fuentes AA, Freeman RA. 2011. A Successful Plan for Faculty Development that has a Lasting Impact. In *2011 ASEE Annual Conference & Exposition*, pp. 22.113.1–22.113.15.

Dedrick RM, Guerrero-Bustamante CA, Garlena RA, Russell DA, Ford K, Harris K, Gilmour KC, Soothill J, Jacobs-Sera D, Schooley RT, et al. 2019. Engineered bacteriophages for treatment of a patient with a disseminated drug-resistant Mycobacterium abscessus. *Nat Med* **25**: 730–733.

Dinsdale E, Elgin SCR, Grandgenett N, Morgan W, Rosenwald A, Tapprich W, Triplett EW, Pauley MA. 2015. NIBLSE: A Network for Integrating Bioinformatics into Life Sciences Education. *CBE Life Sci Educ* **14**: le3.

Elgin SCR, Hays S, Mingo V, Shaffer CD, Williams J. 2021. Building Back More Equitable STEM Education: Teach Science by Engaging Students in Doing Science. *bioRxiv* 2021.06.01.446616. https://www.biorxiv.org/content/10.1101/2021.06.01.446616v1 (Accessed December 17, 2021).

Fuller JA, Luckey S, Odean R, Lang SN. 2021. Creating a diverse, inclusive, and equitable learning environment to support children of color's early introductions to STEM. *Transl Issues Psychol Sci* **7**: 473–486.

Garrison NA, Hudson M, Ballantyne LL, Garba I, Martinez A, Taualii M, Arbour L, Caron NR, Rainie SC. 2019. Genomic Research Through an Indigenous Lens: Understanding the Expectations. *Annu Rev Genomics Hum Genet* **20**: 495–517.

Gin LE, Clark CE, Elliott DB, Roderick TB, Scott RA, Arellano D, Ramirez D, Vargas C, Velarde K, Aeschliman A, et al. 2021. An Exploration across Institution Types of Undergraduate Life Sciences Student Decisions to Stay in or Leave an Academic-Year Research Experience. *CBE Life Sci Educ* **20**: ar47.

Ginther DK, Schaffer WT, Schnell J, Masimore B, Liu F, Haak LL, Kington R. 2011. Race, ethnicity, and NIH research awards. *Science* **333**: 1015–1019.

Glessner K, Rockinson-Szapkiw AJ, Lopez ML. 2017. "yes, I can": Testing an intervention to increase middle school students' college and career self-efficacy. *Career Dev Q* **65**: 315–325.

Goff SA, Vaughn M, McKay S, Lyons E, Stapleton AE, Gessler D, Matasci N, Wang L, Hanlon M, Lenards A, et al. 2011. The iPlant Collaborative: Cyberinfrastructure for Plant Biology. *Front Plant Sci* **2**: 34.

Goodwin S, McPherson JD, McCombie WR. 2016. Coming of age: ten years of next-generation sequencing technologies. *Nat Rev Genet* **17**: 333–351.

Hanauer DI, Graham MJ, SEA-PHAGES, Betancur L, Bobrownicki A, Cresawn SG, Garlena RA, Jacobs-Sera D, Kaufmann N, Pope WH, et al. 2017. An inclusive Research Education Community (iREC): Impact of the SEA-PHAGES program on research outcomes and student learning. *Proc Natl Acad Sci U S A* **114**: 13531–13536.

Hemming J, Eide K, Harwood E, Ali R, Zhu Z, Cutler J, National Research Mentoring Network Coaching Group Directors. 2019. Exploring Professional Development for New Investigators Underrepresented in the Federally Funded Biomedical Research Workforce. *Ethn Dis* **29**: 123–128.

Hoffmann AA, Weeks AR, Sgrò CM. 2021. Opportunities and challenges in assessing climate change vulnerability through genomics. *Cell* **184**: 1420–1425.

Hofstra B, Kulkarni VV, Munoz-Najar Galvez S, He B, Jurafsky D, McFarland DA. 2020. The Diversity-Innovation Paradox in Science. *Proc Natl Acad Sci U S A* **117**: 9284–9291.

Hudson M, Garrison NA, Sterling R, Caron NR, Fox K, Yracheta J, Anderson J, Wilcox P, Arbour L, Brown A, et al. 2020. Rights, interests and expectations: Indigenous perspectives on unrestricted access to genomic data. *Nat Rev Genet* **21**: 377–384.

International Human Genome Sequencing Consortium. 2001. Initial sequencing and analysis of the human genome. *Nature* **409**: 860–921.

Jalili V, Afgan E, Gu Q, Clements D, Blankenberg D, Goecks J, Taylor J, Nekrutenko A. 2020. The Galaxy platform for accessible, reproducible and collaborative biomedical analyses: 2020 update. *Nucleic Acids Res* **48**: W395–W402.

Jarvis T, Weiman S, Johnson D. 2020. Reimagining scientific conferences during the pandemic and beyond. *Sci Adv* **6**. http://dx.doi.org/10.1126/sciadv.abe5815.

Jayabalan M, Caballero ME, Cordero AD, White BM, Asalone KC, Moore MM, Irabor EG, Watkins SE, Walters-Conte KB, Taraboletti A, et al. 2021. Unrealized potential from smaller institutions: Four strategies for advancing STEM diversity. *Cell* **184**: 5845–5850.

Jensen AJ, Bombaci SP, Gigliotti LC, Harris SN, Marneweck CJ, Muthersbaugh MS, Newman BA, Rodriguez SL, Saldo EA, Shute KE, et al. 2021. Attracting Diverse Students to Field Experiences Requires Adequate Pay, Flexibility, and Inclusion. *Bioscience* **71**: 757–770.

Jordan TC, Burnett SH, Carson S, Caruso SM, Clase K, DeJong RJ, Dennehy JJ, Denver DR, Dunbar D, Elgin SCR, et al. 2014. A broadly implementable research course in phage discovery and genomics for first-year undergraduate students. *MBio* **5**: e01051–13.

Khoury MJ, Armstrong GL, Bunnell RE, Cyril J, Iademarco MF. 2020. The intersection of genomics and big data with public health: Opportunities for precision public health. *PLoS Med* **17**: e1003373.

Kokotsaki D, Menzies V, Wiggins A. 2016. Project-based learning: A review of the literature. *Improving Schools* **19**: 267–277.

Krim JS, Coté LE, Schwartz RS, Stone EM, Cleeves JJ, Barry KJ, Burgess W, Buxner SR, Gerton JM, Horvath L, et al. 2019. Models and Impacts of Science Research Experiences: A Review of the Literature of CUREs, UREs, and TREs. *CBE Life Sci Educ* **18**: ar65.

Kross S, Peng RD, Caffo BS, Gooding I, Leek JT. 2020. The democratization of data science education. *Am Stat* **74**: 1–7.

Lee J. 2019. Microbiome CRE Wiki. *GitHub*. https://github.com/joslynnlee/qiime2-workflow-cyverse/wiki/For-Faculty:-Teaching-Instructions (Accessed April 14, 2022).

Lindberg RSN, Laine TH, Haaranen L. 2019. Gamifying programming education in K-12: A review of programming curricula in seven countries and programming games. *Br J Educ Technol* **50**: 1979–1995.

Li X, Carroll CD. 2007. Characteristics of Minority-Serving Institutions and Minority Undergraduates Enrolled in These Institutions: Postsecondary Education Descriptive Analysis Report (NCES 2008-156). *National Center for Education Statistics*. https://eric.ed.gov/?id=ED499114.

Lopatto D, Hauser C, Jones CJ, Paetkau D, Chandrasekaran V, Dunbar D, MacKinnon C, Stamm J, Alvarez C, Barnard D, et al. 2014. A central support system can facilitate implementation and sustainability of a Classroom-based Undergraduate Research Experience (CURE) in Genomics. *CBE Life Sci Educ* **13**: 711–723.

Manolio TA, Rowley R, Williams MS, Roden D, Ginsburg GS, Bult C, Chisholm RL, Deverka PA, McLeod HL, Mensah GA, et al. 2019. Opportunities, resources, and techniques for implementing genomics in clinical care. *Lancet* **394**: 511–520.

Mapes BM, Foster CS, Kusnoor SV, Epelbaum MI, AuYoung M, Jenkins G, Lopez-Class M, Richardson-Heron D, Elmi A, Surkan K, et al. 2020. Diversity and inclusion for the All of Us research program: A scoping review. *PLoS One* **15**: e0234962.

McCormick AC, Zhao C-M. 2005. Rethinking and reframing the carnegie classification. *Change: The Magazine of Higher Learning* **37**: 51–57.

Merchant N, Lyons E, Goff S, Vaughn M, Ware D, Micklos D, Antin P. 2016. The iPlant Collaborative: Cyberinfrastructure for Enabling Data to Discovery for the Life Sciences. *PLoS Biol* **14**: e1002342.

Merisotis JP, McCarthy K. 2005. Retention and student success at minority-serving institutions. *New Dir Inst Res* **2005**: 45–58.

Morgan PL, Farkas G, Hillemeier MM, Maczuga S. 2016. Science Achievement Gaps Begin Very Early, Persist, and Are Largely Explained by Modifiable Factors. *Educational Researcher* **45**: 18–35.

National Center for Biotechnology Information. 2021. SRA Growth. *Sequence Read Archive (SRA)*. https://www.ncbi.nlm.nih.gov/sra/docs/sragrowth/ (Accessed December 7, 2021).

National Institutes of Health. 2021a. Bridges to the Baccalaureate Research Training Program (T34). *Bridges to the Baccalaureate Research Training Program (T34)*. https://grants.nih.gov/grants/guide/pa-files/PAR-19-299.html (Accessed December 17, 2021).

National Institutes of Health. 2021b. IDeA Networks of Biomedical Research Excellence. *IDeA Networks of Biomedical Research Excellence*. https://www.nigms.nih.gov/research/drcb/IDeA/Pages/INBRE.aspx (Accessed 2021).

National Institutes of Health. 2021c. NHGRI creates Office of Training, Diversity and Health Equity. *NHGRI creates Office of Training, Diversity and Health Equity*. https://www.genome.gov/news/news-release/NHGRI-creates-office-of-training-diversity-and-health-equity (Accessed December 8, 2021).

National Institutes of Health. 2021d. NIH Research Enhancement Award (R15). *Grants & Funding*. https://grants.nih.gov/grants/funding/r15.htm (Accessed December 7, 2021).

National Institutes of Health. 2019. Notice of NIH's Interest in Diversity. *Grants & Funding*. https://grants.nih.gov/grants/guide/notice-files/NOT-OD-20-031.html (Accessed December 7, 2021).

National Science Foundation. 2021. Advancing Innovation and Impact in Undergraduate STEM Education at Two-year Institutions of Higher Education. *Advancing Innovation and Impact in Undergraduate STEM Education at Two-year Institutions of Higher Education*. https://beta.nsf.gov/funding/opportunities/advancing-innovation-and-impact-undergraduate-stem-education -two-year (Accessed October 6, 2021).

National Science Foundation. 2017a. Centers of Research Excellence in Science and Technology. *Centers of Research Excellence in Science and Technology*. https://beta.nsf.gov/funding/opportunities/centers-research-excellence-science-and-technology (Accessed 2021).

National Science Foundation. 2020. Historically Black Colleges and Universities - Undergraduate Program (HBCU-UP). *Historically Black Colleges and Universities - Undergraduate Program (HBCU-UP)*. https://beta.nsf.gov/funding/opportunities/historically-black-colleges-and-universities-undergraduate-program-hbcu (Accessed January 9, 2022).

National Science Foundation. 2019a. *NSF: Women, Minorities, and Persons with Disabilities in Science and Engineering: 2019*. National Center for Science and Engineering Statistics https://ncses.nsf.gov/pubs/nsf19304/data (Accessed December 6, 2021).

National Science Foundation. 2017b. Research Coordination Networks. *Research Coordination Networks*. https://beta.nsf.gov/funding/opportunities/research-coordination-networks (Accessed December 7, 2021).

National Science Foundation. 2019b. Research Experiences for Undergraduates (REU). https://beta.nsf.gov/funding/opportunities/research-experiences-undergraduates-reu (Accessed December 7, 2021).

Navale V, Bourne PE. 2018. Cloud computing applications for biomedical science: A perspective. *PLoS Comput Biol* **14**: e1006144.

Nguyen T-H, Lundy-Wagner V, Samayoa A, Gasman M. 2015. On their own terms: Two-year minority serving institutions. https://repository.upenn.edu/cgi/viewcontent.cgi?article=1389&context=gse_pubs.

Rawlings-Goss R. 2019. *Data science careers, training, and hiring: A comprehensive guide to the data ecosystem: How to build a successful data science career, program, or unit*. 1st ed. Springer Nature, Cham, Switzerland.

Robbins SM, Daulton CR, Hurle B, Easter C. 2021. The NHGRI Short Course in Genomics: energizing genetics and genomics education in classrooms through direct engagement between educators and scientists. *Genet Med* **23**: 222–229.

Rowan College at Burlington County. 2021. 3 + 1 Program at Rowan College. *3 + 1 Program at Rowan College*. https://www.rcbc.edu/rowan/3plus1 (Accessed 2021).

Ryder EF, Morgan WR, Sierk M, Donovan SS, Robertson SD, Orndorf HC, Rosenwald AG, Triplett EW, Dinsdale E, Pauley MA, et al. 2020. Incubators: Building community networks and developing open educational resources to integrate bioinformatics into life science education. *Biochem Mol Biol Educ* **48**: 381–390.

Savonen C, Wright C, Hoffman AM, Muschelli J, Cox K, Tan FJ, Leek JT. 2022. Open-source Tools for Training Resources -- OTTR. https://arxiv.org/abs/2203.07083.

Schatz MC, Philippakis AA, Afgan E, Banks E, Carey VJ, Carroll RJ, Culotti A, Ellrott K, Goecks J, Grossman RL, et al. 2021. Inverting the model of genomics data sharing with the NHGRI Genomic Data Science Analysis, Visualization, and Informatics Lab-space (AnVIL). *bioRxiv* 2021.04.22.436044. https://www.biorxiv.org/content/10.1101/2021.04.22.436044v1 (Accessed May 5, 2021).

Sirugo G, Williams SM, Tishkoff SA. 2019. The Missing Diversity in Human Genetic Studies. *Cell* **177**: 26–31.

Stich AE, Reeves TD. 2017. Massive open online courses and underserved students in the United States. *Internet High Educ* **32**: 58–71.

Sweeder R, Kursav MN, Valles S. 2021. A Cohort Scholarship Program that Reduces Inequities in STEM retention. *JSTEM* **22**. https://jstem.org/jstem/index.php/JSTEM/article/view/2456 (Accessed January 9, 2022).

Texas Education Agency. 2020. Early College High School (ECHS). *Early College High School (ECHS)*. https://tea.texas.gov/academics/college-career-and-military-prep/early-college-high-school-echs (Accessed December 17, 2021).

Virginia Community Colleges. 2020. High School Dual Enrollment. *High School Dual Enrollment*. https://www.vccs.edu/high-school-dual-enrollment/ (Accessed December 17, 2021).

Whittaker JA, Montgomery BL, Martinez Acosta VG. 2015. Retention of underrepresented minority faculty: Strategic initiatives for institutional value proposition based on perspectives from a range of academic institutions. *J Undergrad Neurosci Educ* **13**: A136–45.

Wiley K, Dixon BE, Grannis SJ, Menachemi N. 2020. Underrepresented racial minorities in biomedical informatics doctoral programs: graduation trends and academic placement (2002-2017). *J Am Med Inform Assoc* **27**: 1641–1647.

Williams JJ, Drew JC, Galindo-Gonzalez S, Robic S, Dinsdale E, Morgan WR, Triplett EW, Burnette JM 3rd, Donovan SS, Fowlks ER, et al. 2019. Barriers to integration of bioinformatics into undergraduate life sciences education: A national study of US life sciences faculty uncover significant barriers to integrating bioinformatics into undergraduate instruction. *PLoS One* **14**: e0224288.

Wilson Sayres MA, Hauser C, Sierk M, Robic S, Rosenwald AG, Smith TM, Triplett EW, Williams JJ, Dinsdale E, Morgan WR, et al. 2018. Bioinformatics core competencies for undergraduate life sciences education. *PLoS One* **13**: e0196878.

Winkle-Wagner R, McCoy DL. 2018. Feeling like an "Alien" or "Family"? Comparing students and faculty experiences of diversity in STEM disciplines at a PWI and an HBCU. *Race Ethnicity and Education* **21**: 593–606.

Xie X, Lewis T-J, Green N, Wang Z. 2021. Phylogenetic network analysis revealed the recombinant origin of the SARS-CoV-2 VOC202012/01 (B.1.1.7) variant first discovered in U.K. *bioRxiv* 2021.06.24.449840. https://www.biorxiv.org/content/10.1101/2021.06.24.449840v2 (Accessed January 9, 2022).

Yarmohammadian MH, Davidson P, Yeh CH. 2018. Sabbatical as a part of the academic excellence journey: A narrative qualitative study. *J Educ Health Promot* **7**: 119.

Zhan YA, Wray CG, Namburi S, Glantz ST, Laubenbacher R, Chuang JH. 2019. Fostering bioinformatics education through skill development of professors: Big Genomic Data Skills Training for Professors. *PLoS Comput Biol* **15**: e1007026.

Zook M, Barocas S, Boyd D, Crawford K, Keller E, Gangadharan SP, Goodman A, Hollander R, Koenig BA, Metcalf J, et al. 2017. Ten simple rules for responsible big data research. *PLoS Comput Biol* **13**: e1005399.

**Supplemental information to accompany:**

*Diversifying the Genomic Data Science Research Community*

**Text S1:** Faculty and staff at UIs (Underserved Institutions) possess unique strengths and have experience with the specific interests, needs, challenges, and concerns of the populations they serve. We briefly describe these unique attributes below.

- **Community Colleges (CCs)** have smaller class sizes at the freshman and sophomore levels and provide unique student-centered support (Holmberg et al. 2021). Tuition at CCs is extremely affordable compared to public universities (on average only one third the cost), and flexible class schedules accommodate working students. CCs tend to be more agile in responding to the needs of the community and provide support to traditional and non-traditional students alike, often supplying more personalized attention. CCs promote a culture of inclusion and innovation while making higher education more accessible to individuals across socioeconomic groups. Large teaching loads (e.g., 4 lectures and 3 labs per semester) with no release time are typical for CC faculty. Research laboratories, equipment, and funds for laboratory supplies are limited or completely absent, making experimental work extremely challenging, but data science offers an exciting opportunity.

- **Historically Black Colleges and Universities (HBCUs)** are incredibly diverse and include institutions that focus on liberal arts, business, professional degrees, workforce development, or cutting-edge research. HBCUs have been key to increasing participation and success of minority students in STEM (Gasman and Nguyen 2016) and provide an affordable, supportive learning environment for minority student achievement (Harper 2019). HBCUs serve as pillars of Black and African American history, art, culture, and politics, attracting an extremely diverse student body and faculty community. However, the legacy of slavery, segregation, and systemic social discrimination means that HBCUs tend to have financial challenges (Harper 2019). Some faculty members at HBCUs are developing new federally funded education and research programs to provide students training in genomics and data sciences. However, the long-term success of these programs requires wider acceptance by HBCU faculty colleagues and support from administrations in the form of teaching release time, compensation, technical personnel, and adequate genomics and data science research facilities.

- ***Hispanic-Serving Institutions (HSIs)*** provide an essential resource making education accessible to the nation's growing population of Hispanic Americans. HSIs enroll two-thirds of all Hispanic undergraduates and are among the top institutions with respect to the Social Mobility Index (SMI) (Hispanic Association of Colleges and Universities 2021). Students choose HSIs for their exceptional affordability and proximity to home and family (Cuellar 2019). Like HBCUs, HSIs are under-resourced. These diverse institutions receive only 68 cents for every federal funding dollar granted to all other institutions per student annually (Hispanic Association of Colleges and Universities 2021)).

- ***Tribal Colleges and Universities (TCUs)*** provide culturally relevant, geographically accessible, and affordable higher education. Indigenous ways of thinking bring diverse perspectives and solutions to health and environmental questions. TCUs are centered around the communities they serve, forming bridges to meaningful community engagement. Outside collaborations with TCUs must be flexible, reciprocal, and culturally sensitive. TCUs and affiliated tribal communities require engagement through every step of the research and data sharing process (Hudson et al. 2020). Indigenous scientists are wary of "helicopter research"; collaborations should create meaningful educational and training materials that can lead to co-publication and co-presentation (Guglielmi 2019; Fox 2020). Activities should be sustainable and empowering for TCU faculty. Academic researchers engaged in tribal projects should become familiar with tribal sovereignty and ethics and informed consent (Garrison et al. 2019) as well as community education around genetics (Claw et al. 2021). Collaborative research projects and training opportunities may focus on tribal health or environmental priorities. TCUs are continuing to work on their technology and internet infrastructure to increase access in remote areas.

**Table S1:** Organizations supporting the needs of underserved subpopulations have emerged in recent years. These groups are vital to developing a sense of belonging and support system. Some examples of these affinity groups include those listed here.

| Organization | Website |
|---|---|
| Center for First-generation Student Success | https://firstgen.naspa.org/ |
| Student Veterans Research Network | https://www.svrn.org/ |
| Society for Advancement of Chicanos/Hispanics and Native Americans in Science | https://www.sacnas.org/ |
| American Indian Science and Engineering Society | https://www.aises.org/ |
| Native BioData Consortium | https://indigidata.nativebio.org/ |
| Annual Biomedical Research Conference for Minority Students | https://abrcms.org/ |
| National Foster Youth Institute | https://nfyi.org/issues/higher-education/ |
| From Prison Cells to PhD | https://www.fromprisoncellstophd.org/ |

**Table S2***: Cloud-based computing resources. Note that other tools, such as JupyterHub (https://jupyter.org/hub#deploy-a-jupyterhub) and Binder (https://mybinder.org/), can also make it easier to deploy notebooks for multiple users on commercially available clouds using Kubernetes. The Galaxy Server Directory (https://galaxyproject.org/use/) also includes a comprehensive list of more cloud providers providing access to Galaxy. Combinations of open source software (e.g., Docker and R, https://www.docker.com/blog/docker-higher-education-tools-resources-teachers/) can also provide an alternative for modular teaching needs. Online resources change frequently; this information was collected on Dec 7, 2021.*

| Resource | Interface | Costs | Strengths / Weaknesses |
|---|---|---|---|
| *Project Specific* | | | |
| AnVIL / Terra https://anvilproject.org/learn | Terra portal for launching cloud environments with workflows, Jupyter Notebook, RStudio, Galaxy, and Command Line Interface; hosts open and controlled NHGRI datasets | **Compute** starting at $0.06 per hr; possible to obtain $300 in credits through Google Cloud **Persistent disk** $4.00 per month per 100GB | **Strengths:** Collaborative work on sensitive/protected and NHGRI datasets, many options for different experience levels, extremely scalable **Weaknesses:** Lacks a free tier for beginners |
| All of Us https://www.researchallofus.org/ | Terra portal that includes analysis workspaces and Jupyter Notebook; hosts open and protected tier for All of Us Research Program participant data | **Compute** starting at $0.06 per hr; possible to obtain $300 in credits through Google Cloud **Persistent disk** $4.00 per month per 100GB | **Strengths:** Access and ability to build datasets, built in analysis spaces **Weaknesses:** Lacks a free tier for beginners |
| BioData Catalyst https://biodatacatalyst.nhlbi.nih.gov | Terra and Seven Bridges portal that includes analysis workflows and RStudio and Jupyter Notebook; hosts protected NHLBI datasets | **Compute** costs vary depending on AWS instance used; possible to obtain $500 in credits **Persistent disk** $2.10 per month per 100GB through Seven Bridges | **Strengths:** Access to datasets, built in workflows/pipelines **Weaknesses:** Lacks a free tier for beginners |
| Cancer Research Data Commons https://datacommons.cancer.gov/ | Terra, Seven Bridges, and ISB-CGC portal that includes analysis workflows, genomic tools, RStudio, and Jupyter Notebook; hosts open and controlled datasets from NCI programs and key external cancer programs | **Compute** and **Persistent disk** costs vary depending on platform used, possible to obtain $300 in credits through Google Cloud | **Strengths:** Access to datasets, built in workflows/pipelines, flexible with many options for different experience levels **Weaknesses:** Lacks a free tier for beginners |
| Kids First DRC https://kidsfirstdrc.org | Storage, sharing, and analysis portal supported by Cavatica / Seven Bridges with workflows and genomics apps; hosts open and controlled datasets | **Compute** costs vary depending on AWS instance used; possible to obtain $500 in credits **Persistent disk** $2.10 per month per 100GB through Seven Bridges | **Strengths:** Access to datasets, built in workflows/pipelines **Weaknesses:** Lacks a free tier for beginners, less flexibility |
| UK Biobank https://ukbiobank.dnanexus.com/landing | DNANexus portal for launching cloud environments with workflows, JupyterLab, Command | **Compute** and **Persistent disk** costs vary depending on AWS instance used; users start with | **Strengths:** Access to datasets, flexible with many options for analysis |

| | Line Interface, or custom tools; hosts the UK Biobank dataset | approximately $50 in free compute and storage credit | **Weaknesses:** Learning curve might be steeper compared to other platforms |
|---|---|---|---|
| *Academic* | | | |
| CyVerse https://learning.cyverse.org/ | Launching platform for maintained images, including those with bioInformatics tools, Jupyter Notebook, RStudio, and Galaxy | **Compute** free with short/moderate wait times with subscriptions available **Persistent disk** free up to 100GB, starting at $100 per TB per year for >100GB. | **Strengths:** Affordable, many options for different experience levels **Weaknesses:** Requires permission to access resources |
| Galaxy https://training.galaxyproject.org/ | Point-and-click for commonly used genomics tools | **Compute** free tier limited to 6 compute jobs for registered users, 1 for unregistered users (short/moderate wait times) **Persistent disk** free tier limited to 250GB for registered users, 5GB for unregistered users | **Strengths:** No programming experience needed, powerful workflows, free for basic analyses **Weaknesses:** Wait times/limits, limited to tools available |
| GenePattern https://www.genepattern.org/user-guide (Reich et al. 2006) | Point-and-click for commonly used genomics tools, plus Jupyter Notebook | **Compute** free tier with short/moderate wait times **Persistent disk** free public server limited to 30GB. | **Strengths:** No programming experience needed, free for basic analyses **Weaknesses:** limited to tools available, less active user community |
| Jetstream https://jetstream-cloud.org/documentation-training/index.html (Stewart et al. 2015) | Launching platform for maintained images, including those with bioInformatics tools, R, Galaxy, and more | **Compute** free tier up to 50,000 Virtual CPU Hours **Persistent disk** allocations provided on a per-project basis; Allocation requests for education and research available (https://docs.jetstream-cloud.org/faq/alloc/). | **Strengths:** Affordable, many options for different experience levels **Weaknesses:** Requires permission to access resources |
| KBase https://www.kbase.us/learn/ (Arkin et al. 2018) | Point-and-click for commonly used genomics tools, plus Jupyter Notebook | **Compute** free tier with short/moderate wait times. **Persistent disk** has no strict storage limits | **Strengths:** Easy start on free tier, No programming experience needed **Weaknesses:** Less flexible, limited to tools available |
| SciServer https://www.sciserver.org/support/ (Taghizadeh-Popp et al. 2020) | Launching platform for maintained images, including those with bioInformatics tools, R, python, and more; hosts datasets | **Compute** free tier with short/moderate wait times **Persistent disk** free up to 10GB permanent storage; temporary storage at 1TB+ | **Strengths:** Access to large datasets, flexible with many options for analysis; great option for collaborating with physicists and other multidisciplinary team members **Weaknesses:** Learning curve might be steeper compared to other platforms |
| *Commercial* | | | |
| Illumina BaseSpace https://basespace.illumina.com/ | Point-and-click for commonly used genomics tools | **Compute** with limited free basic tier; credits can be purchased **Persistent disk** free tier limited to 1TB; credits can be purchased | **Strengths:** Built-in workflows, collaborative work on sensitive/protected datasets **Weaknesses:** Less flexible, more expensive than others |
| Google Colaboratory https://colab.research.google.com/ | Jupyter notebook (Python) | **Compute** free tier resources are limited, usage limits fluctuate **Persistent disk** free up to 5GB | **Strengths:** Easy start on free tier, Quickly code live with other users |
| | | | **Weaknesses:** Connection and RAM not guaranteed, limited languages/tools |
| RStudio Cloud<br>https://rstudio.cloud/ | RStudio accessed through browser | **Compute** free tier up to 25 hours, paid tier $0.10 per hour<br>**Persistent disk** free tier limited to 20GB per project | **Strengths:** Easy start on free tier<br>**Weaknesses:** limited to R based languages/tools |
| Seven Bridges<br>https://www.sevenbridges.com/platform/ | Point-and-click for commonly used genomics tools, RStudio, command line interface | **Compute** and **Persistent disk** prices determined by AWS negotiated price | **Strengths:** Flexible and scalable, with built in genomics tools, relatively affordable<br>**Weaknesses:** No free tier for beginners |

**Table S3:** Learning resources to supplement courses in genomic data science. MOOC: Massive online open course. If needed, subscription or paid services should be covered by funding sources.

| Description | Resource |
|---|---|
| Standalone MOOC / Course for genomic data science | • Coursera Genomic Data Science Specialization (https://www.coursera.org/specializations/genomic-data-science)<br>• Bioinformatics Algorithms (https://www.bioinformaticsalgorithms.org/) |
| Project based learning modules (open source) | • Open Case Studies Project (https://www.opencasestudies.org/) |
| Lesson planning network; modules focused on bioinformatics and data science (open source) | • Quantitative Undergraduate Biology Education and Synthesis Hub (QUBES, https://qubeshub.org/publications/browse) (Donovan et al. 2015) |
| Modular training for genetics, genomics, and bioinformatics (open source) | • CyVerse tutorials (https://learning.cyverse.org/en/latest/tutorials.html)<br>• Galaxy Training Network (https://training.galaxyproject.org/)<br>• Orchestra (http://app.orchestra.cancerdatasci.org/)<br>• Babraham Institute bioinformatics courses (https://www.bioinformatics.babraham.ac.uk/training.html)<br>• XBio Cell Biology & Genetics (https://explorebiology.org/collections/genetics) |
| Modular training for data science (open source) | • Tidyverse Skills for Data Science in R (https://leanpub.com/tidyverseskillsdatascience)<br>• Learn-R (https://www.learn-r.org/)<br>• Swirl (https://swirlstats.com/)<br>• R for Data Science (https://r4ds.had.co.nz/)<br>• Data Science in Practice (https://datascienceinpractice.github.io/) (Donoghue, Voytek, and Ellis 2021)<br>• The Carpentries (https://carpentries.org/)<br>• SciServer Courseware (https://www.sciserver.org/outreach/)<br>• Python for Biologists (https://www.pythonforbiologists.org/) |
| Modular training (fee- or subscription-based) | • DataQuest (https://www.dataquest.io/)<br>• Codecademy (https://www.codecademy.com)<br>• Data Camp (https://www.datacamp.com/)<br>• SimBio (https://simbio.com/) |
| Modular training for K-12 | • DataNuggets (http://datanuggets.org/) (Schultheis and Kjelvik 2015)<br>• DataSpire (https://dataspire.org)<br>• Oak Ridge Institute for Science and Education (https://orise.orau.gov/resources/k12/lesson-plans.html)<br>• YouCubed (https://www.youcubed.org/) |

**Table S4:** List of courses and topics which could be included in an accredited degree program, based on the St. Mary's University B.S. in Bioinformatics. Prerequisites include General Biology for Majors, General Chemistry, and Calculus I; corequisites include Fundamentals of programming/software development and General Physics. For more information, see

https://catalog.stmarytx.edu/undergraduate/majors-programs/science-engineering-technology/bioinformatics/bioinformatics-bl/#degreeplantext

| Course name | Topics | Learning Objectives |
|---|---|---|
| Introduction to Bioinformatics (w/lab) | Basics of genomics and bioinformatics<br>Human genome and browser<br>Pairwise Sequence alignment<br>Multiple Sequence alignment<br>Sequence Database Search<br>Sequence polymorphism<br>Phylogenetic analysis<br>Gene finding/prediction<br>RNA sequences: prediction and analysis of structures | The objective of this course is to teach how computational techniques can help with solving biological problems. Students will learn to efficiently use multiple genomics and bioinformatics tools, that are freely available, for the analysis of DNA, RNA and protein sequences and structure. No programming skills are necessary for this course. |
| Biostatistics for Life Sciences | How to use R for data analysis.<br>Samples and Populations<br>Linear Regression<br>Comparison of Groups<br>The Normal Distribution<br>Statistical Models, Estimation, and Confidence Intervals<br>Hypothesis Tests<br>Probabilities<br>The Binomial Distribution<br>Logistic Regression<br><br>Survival Analysis | This course will provide the background and application of statistical tools for analyzing different types of data frequently encountered by life scientists. The emphasis will be on the applications of various statistical methodologies on biological data, using the R programming language. |
| Genes, Genomes, and Genomics (w/lab) | From Genes to Genomes: basic molecular biology<br>How to clone a gene<br>Genomic and cDNA Libraries<br>Polymerase Chain Reaction<br>Sequencing a Cloned Gene<br>Analysis of Gene Expression<br>Products from Native and Manipulated Cloned Genes<br>Genomic Analysis<br>Analysis of Genetic Variation<br>Transgenesis | The objective of this course is to teach students the basics of molecular biology. The focus of this course will be on genes, genomes and genomics. This will include strategies to clone a gene and prepare genomic and cDNA libraries. Students will learn the basics of DNA sequencing. They will also learn to analyze gene expression. The course will also focus on analysis of genomes and genetic variation. Students will also learn about transgenesis. |
| Transcriptomics, Proteomics, and Metabolomics (w/lab) | DNA microarray technology Challenges and Future Trends in DNA Microarray Analysis Next Generation Sequencing: New Tools Overview of Quantitative Proteomic Approaches Overview of Protein Microarrays Transcriptome and Metabolome Data Integration Identification of Biomarkers and Biochemical Pathway Visualization Functional Glycomics Analysis: Challenges and Methodologies Applications of Glycan Microarrays to Functional Glycomics Bioinformatic Analysis of Gene Expression Data Transcriptome and Metabolome Data Integration | The objective of this course is to teach students the fundamental aspects of the new instrumental and methodological developments in omics technologies, including those related to genomics, transcriptomics, epigenetics, proteomics and metabolomics. The focus of this course will be on DNA microarray analysis, next-generation sequencing technologies, genome-wide analysis of methylation and histone modifications. Students will learn emerging techniques in proteomics and recent quantitative proteomics approaches. They will also learn the basics of metabolomics and metabolome analysis. The course will also focus on statistical approaches for the analysis of microarray data, the integration of transcriptome and metabolome data and computational approaches for visualization and integration of omics data. |
| Algorithms for Computational Biology with PERL/Python | Linux for bioinformatics Getting started with Perl/Python Individual approaches to programming Representing and Manipulating sequence data Using PERL/Python documentation Motifs and Loops Operating Strings and Arrays Regular expressions Hashes and Data Structures Creating Subroutines | The objective of this course is to teach students the basics of theLinux environment and PERL scripting. Students will learn how to write PERL scripts for solving biological problems. The focus of this course will be on designing algorithms to manipulate and analyze sequence data. This will include programming strategies to store and concatenate DNA sequences. Writing scripts to generate complementary and reverse complementary sequences. Students will learn how to work with files and arrays. They will also learn to generate random numbers and simulate DNA mutations. The course will also focus on hashes and data structures. |
| Programming for Bioinformatics with R | DNA Sequence Statistics Sequence Databases Pairwise Sequence Alignment Multiple Alignment and Phylogenetic trees Computational Gene-finding Comparative Genomics Hidden Markov Models Protein-Protein Interaction Graphs | The objective of this course is to teach students bioinformatics programming techniques using R. The focus of this course will be on development of programs to perform commonly used bioinformatics techniques like pairwise and multiple sequence alignments. Students will learn computational gene-finding and comparative genomic techniques. They will also learn to create phylogenetic trees and protein-protein interaction graphs. Students will also learn about Hidden Markov models. |
| Big Data Concepts | Basic Big Data Concepts<br>Linux for Big Data Analysis<br>Python for Big Data Analysis<br>R for Big Data Analysis<br>Genome-Seq Data Analysis<br>RNA-Seq Data Analysis<br>Microbiome-Seq Data Analysis<br>miRNA-Seq Data Analysis<br>ChIP-Seq Data Analysis<br>Big Data and Drug Discovery | The objective of this course is to teach students the basic big data concepts. The focus of this course will be on big data analysis. Students will learn emerging techniques in proteomics and recent quantitative proteomics approaches. They will also learn the basics of the platforms and programming languages used for big data analysis. The course will also focus on the analysis of Genome-, RNA-, miRNA-, Microbiome- and ChIP-sequencing data. Students will also learn about the usage of big data in drug discovery. |
| Bioinformatics Internship | 120 hrs of documented internship or research. | The objective of this course is to provide an opportunity for students in Bioinformatics major to participate in real-life bioinformatics internship or research. This course will be for seniors and juniors. Emphasis will be placed on commonly used genomics/transcriptomics/proteomics/metabolomics projects and the use of standard operating laboratory/industry procedures. Examples of potential collaborative organizations include medical/health centers, molecular genomics labs/companies, computational biology labs/companies, software development/labs companies and biostatistics labs/companies. |
| Bioinformatics Capstone | **Thesis -**<br>Prior to the end of the semester, write and submit a 25-30 page, double-spaced summary of the research to the capstone instructor. This will include an abstract, introduction, materials and methods, results and discussion and conclusion and references.<br>The thesis should have at least 3 figures with legends and descriptions.<br>**Oral presentation**<br>A 20 minutes' powerpoint presentation consisting of a minimum of 20 slides<br>Oral presentation will be on the same research that is presented in the thesis<br>5 minutes of Question and Answer sessions with peers and faculty members | The objective of this integrative Capstone is to provide students with an opportunity to write a thesis on their research and present it in the form of an oral presentation. This course will promote advanced scientific writing and broad perspectives of issues in current Bioinformatics research. Students will demonstrate their ability to integrate concepts to a practical situation by presenting a thesis on the research they have performed in an industrial or academic setting. The capstone must be taken during the senior year. |

# References

Arkin, Adam P., Robert W. Cottingham, Christopher S. Henry, Nomi L. Harris, Rick L. Stevens, Sergei Maslov, Paramvir Dehal, et al. 2018. "KBase: The United States Department of Energy Systems Biology Knowledgebase." *Nature Biotechnology* 36 (7): 566–69.

Claw, Katrina G., Nicolas Dundas, Michael S. Parrish, Rene L. Begay, Travis L. Teller, Nanibaa' A. Garrison, and Franklin Sage. 2021. "Perspectives on Genetic Research: Results From a Survey of Navajo Community Members." *Frontiers in Genetics* 12 (December): 734529.

Cuellar, Marcela G. 2019. "Creating Hispanic-Serving Institutions (HSIs) and Emerging HSIs: Latina/o College Choice at 4-Year Institutions." *American Journal of Education* 125 (2): 231–58.

Donoghue, Thomas, Bradley Voytek, and Shannon E. Ellis. 2021. "Teaching Creative and Practical Data Science at Scale." *Journal of Statistics and Data Science Education* 29 (sup1): S27–39.

Donovan, Sam, Carrie Diaz Eaton, Stith T. Gower, Kristin P. Jenkins, M. Drew LaMar, Dorothybelle Poli, Robert Sheehy, and Jeremy M. Wojdak. 2015. "QUBES: A Community Focused on Supporting Teaching and Learning in Quantitative Biology." *Letters in Biomathematics* 2 (1): 46–55.

Fox, Keolu. 2020. "The Illusion of Inclusion - The 'All of Us' Research Program and Indigenous Peoples' DNA." *The New England Journal of Medicine* 383 (5): 411–13.

Garrison, Nanibaa' A., Māui Hudson, Leah L. Ballantyne, Ibrahim Garba, Andrew Martinez, Maile Taualii, Laura Arbour, Nadine R. Caron, and Stephanie Carroll Rainie. 2019. "Genomic Research Through an Indigenous Lens: Understanding the Expectations." *Annual Review of Genomics and Human Genetics* 20 (August): 495–517.

Gasman, Marybeth, and Thai-Huy Nguyen. 2016. "Engaging Voices: Methods for Studying STEM Education at Historically Black Colleges and Universities (HBCUs)." *Journal for Multicultural Education* 10 (2): 194–205.

Guglielmi, Giorgia. 2019. "Facing up to Injustice in Genome Science." *Nature* 568 (7752): 290–93.

Harper, Brian E. 2019. "African American Access to Higher Education: The Evolving Role of Historically Black Colleges and Universities." *American Academic* 3 (January). http://hdl.handle.net/10919/86971.

Hispanic Association of Colleges and Universities. 2021. "2021 Hispanic Higher Education and HSIs Facts." Hispanic Association of Colleges and Universities. April 6, 2021. https://www.hacu.net/hacu/HSI_Fact_Sheet.asp.

Holmberg, Tara Jo, Sharon Gusky, Stacey Kiser, Vedham Karpakakunjaram, Heather Seitz, Linnea Fletcher, Lindsey Fields, Apryl Nenortas, Andrew Corless, and Katrina Marcos. 2021. "Biology Educators, Professional Societies, and Practitioner Networks within Community Colleges." *New Directions for Community Colleges* 2021 (194): 15–28.

Hudson, Maui, Nanibaa' A. Garrison, Rogena Sterling, Nadine R. Caron, Keolu Fox, Joseph Yracheta, Jane Anderson, et al. 2020. "Rights, Interests and Expectations: Indigenous Perspectives on Unrestricted Access to Genomic Data." *Nature Reviews. Genetics* 21 (6): 377–84.

Reich, Michael, Ted Liefeld, Joshua Gould, Jim Lerner, Pablo Tamayo, and Jill P. Mesirov. 2006. "GenePattern 2.0." *Nature Genetics* 38 (5): 500–501.

Schultheis, Elizabeth H., and Melissa K. Kjelvik. 2015. "Data Nuggets." *The American Biology Teacher* 77 (1): 19–29.

Stewart, Craig A., George Turner, Matthew Vaughn, Niall I. Gaffney, Timothy M. Cockerill, Ian Foster, David Hancock, et al. 2015. "Jetstream." In *Proceedings of the 2015 XSEDE Conference on Scientific Advancements Enabled by Enhanced Cyberinfrastructure - XSEDE '15*. New York, New York, USA: ACM Press. https://doi.org/10.1145/2792745.2792774.

Taghizadeh-Popp, M., J. W. Kim, G. Lemson, D. Medvedev, M. J. Raddick, A. S. Szalay, A. R. Thakar, et al. 2020. "SciServer: A Science Platform for Astronomy and beyond." *Astronomy and Computing* 33 (100412): 100412.